\documentclass[pra,10pt,twocolumn,floatfix,groupedaddress]{revtex4-1} 

\usepackage[dvips]{graphicx}
\usepackage{subfigure}
\usepackage{amsmath}
\usepackage{amssymb}
\usepackage{bm}

\graphicspath{{./}}
\def\Eq{Eq.~}

\def\Fig{Fig.~}

\def\Ref{Ref.~}
\def\Refs{Refs.~}
\def\Sec{Sec.~}

\def\be{\begin{equation}}
\def\ee{\end{equation}}
\def\bea{\begin{eqnarray}}
\def\eea{\end{eqnarray}}

\def\ie{i.e.~}

\newcommand{\ket}[1]{\left| #1 \right\rangle}

\newcommand{\refeqn}[1]{(\ref{#1})}

\begin{document}

\title{Measuring the atomic recoil frequency using a perturbative\\grating-echo atom interferometer}

\author{B. Barrett}
\altaffiliation[Current address: ]{Laboratoire Photonique Num\'{e}rique et Nanosciences, Universit\'{e} Bordeaux 1, IOGS and CNRS, 351 cours de la Lib\'{e}ration, 33405 Talence, France}
\author{A. Carew}
\author{S. Beattie}
\altaffiliation[Current address: ]{Department of Physics, University of Toronto, Toronto, Ontario M5S 1A7, Canada}
\author{A. Kumarakrishnan}
\affiliation{Department of Physics \& Astronomy, York University, Toronto, Ontario M3J 1P3, Canada}

\date{\today}


\begin{abstract}
We describe progress toward a precise measurement of the recoil energy of an atom measured using a perturbative grating-echo atom interferometer (AI) that involves three standing-wave (sw) pulses. With this technique, a perturbing sw pulse is used to shift the phase of excited momentum states---producing a modulation in the contrast of the interference pattern. The signal exhibits narrow fringes that revive periodically at twice the two-photon recoil frequency, $2\omega_q$, as a function of the onset time of the pulse. Experiments are performed using samples of laser-cooled rubidium atoms with temperatures $\lesssim 5$ $\mu$K in a non-magnetic apparatus. We demonstrate a measurement of $\omega_q$ with a statistical uncertainty of 37 parts per $10^9$ (ppb) on a time scale of $\sim 45$ ms in 14 hours. Further statistical improvements are anticipated by extending this time scale and narrowing the signal fringe width. However, the total systematic uncertainty is estimated to be $\sim 6$ parts per $10^6$ (ppm). We describe methods of reducing these systematic errors.
\end{abstract}

\maketitle

\section{Introduction}

There is an ongoing, international effort to develop precise, independent techniques for measuring the atomic fine structure constant, $\alpha$---a dimensionless parameter that quantifies the strength of the electromagnetic force. These measurements can be used to stringently test the theory of quantum electrodynamics (QED). Historically, two types of determinations of $\alpha$ have been carried out: (i) those that use other precisely measured quantities to determine $\alpha$ through challenging QED calculations \cite{Aoyama-PRL-2007, Pachucki-PRA-2012}, and (ii) those that are independent of QED. The latter depend on only the quantities appearing in the definition $\alpha \equiv e^2/2 \epsilon_0 h c$, where $e$ is the elementary charge, $\epsilon_0$ is the vacuum permittivity, $h$ is Planck's constant and $c$ is the speed of light. Some examples of $\alpha$ determinations that require QED are the measurements of the anomalous magnetic moment of the electron \cite{Hanneke-PRL-2008}, and the fine structure intervals of helium \cite{Smiciklas-PRL-2010}. The most precise examples of QED-independent determinations are those based on measurements of the von Klitzing constant, $R_{\rm{K}} = h/e^2$, using the quantum-Hall effect \cite{Jeffery-IEEE-1997, Small-Metrologia-1997}, and the ratio $h/M$ using (i) Bloch oscillations in cold atoms \cite{Clade-PRL-2006} and (ii) atom interferometric techniques \cite{Weiss-PRL-1993, Wicht-PhysScr-2002, Cadoret-PRL-2008, Bouchendira-PRL-2011}. Within these examples, atom interferometry has emerged as a powerful tool because of its inherently high sensitivity to $h/M$, which can be related to $\alpha$ according to
\be
  \label{eqn:alpha2}
  \alpha^2
  = \frac{2 R_{\infty}}{c} \frac{h}{m_e}
  = \frac{2 R_{\infty}}{c} \left(\frac{M}{m_e}\right) \left(\frac{h}{M}\right).
\ee
Here, $R_{\infty}$ is the Rydberg constant, $m_e$ is the electron mass, and $M$ is the mass of the test atom. Since $R_{\infty}$ is known to 5 parts in $10^{12}$, and the mass ratio $M/m_e$ is typically known to a few parts in $10^{10}$ \cite{Mohr-arXiv-2012}, the quantity that limits the precision of a determination of $\alpha$ using \Eq \refeqn{eqn:alpha2} is the ratio $h/M$. The most precise measurement of this ratio was recently carried out in $^{87}$Rb, where $h/M(^{87}\mbox{Rb})$ was determined to $1.2$ ppb \cite{Bouchendira-PRL-2011}. Coupled with the most precise measurement of the electron anomaly, $a_e$ \cite{Hanneke-PRL-2008}, this work demonstrated the importance of hadronic and weak-interaction terms in the series expansion of $a_e$ in powers of $\alpha$. Other interferometric techniques that have demonstrated high sensitivity to $h/M$ include \Refs \cite{Weitz-PRL-1994, Gupta-PRL-2002, Muller-ApplPhysB-2006, Chiow-PRL-2009, Chiow-PRL-2011}.

In recent years, the grating-echo AI has emerged as a candidate for precise measurements of the two-photon recoil frequency, $\omega_q = \hbar q^2/2M$, where $\hbar q = 2\hbar k$ is the two-photon momentum and $k = 2\pi/\lambda$ is the wave number of the excitation light \cite{Cahn-PRL-1997}. The appeal of this AI lies in its reduced experimental complexity compared to the more broadly used, Raman-transition-based interferometers. Specifically, the grating-echo AI does not require internal state or velocity selection, and it utilizes only one laser frequency. Also, since this interferometer uses a single hyperfine ground state, it has reduced sensitivity to common systematic effects such as the ac Stark and Zeeman shifts. Low-frequency phase noise in the sw excitation beam due to mirror vibrations is also a negligible concern for three main reasons: (i) the excitation pulses are short-lived ($\lesssim 1$ $\mu$s), (ii) the phase introduced by each sw pulse is common to all excited momentum states, and (iii) the interference is probed using intensity detection, which is insensitive to the phase of the back-scattered light.

\begin{figure*}[!tb]
  \centering
  \subfigure{\includegraphics[width=0.66\textwidth]{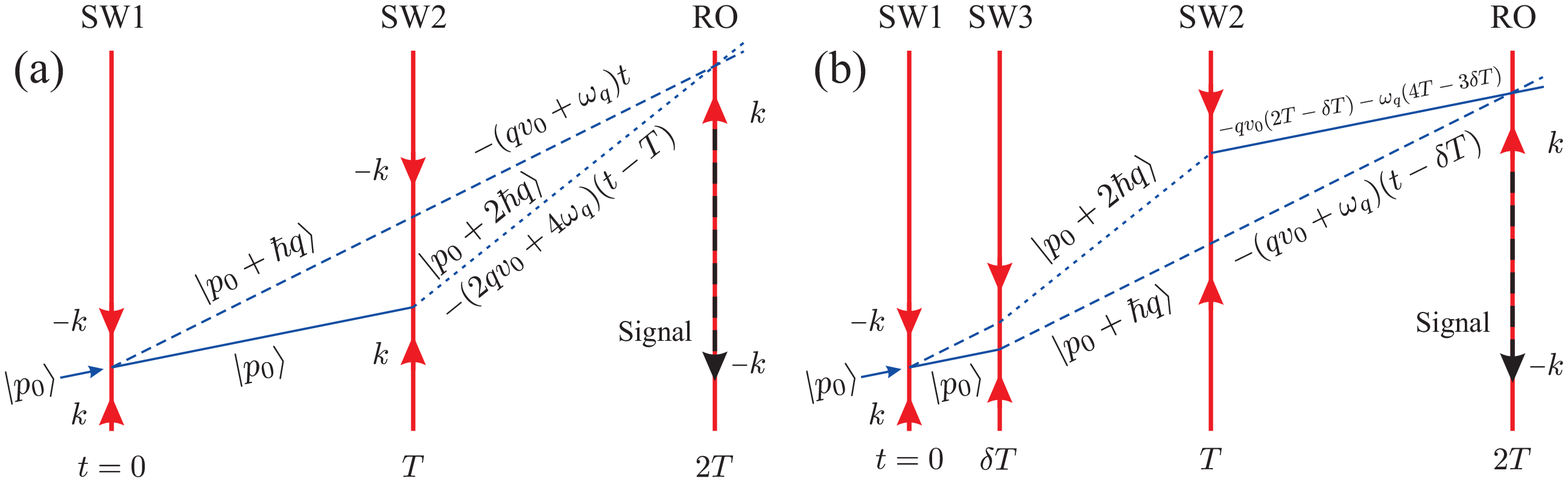}}
  \hspace{0.1cm}
  \subfigure{\includegraphics[width=0.30\textwidth]{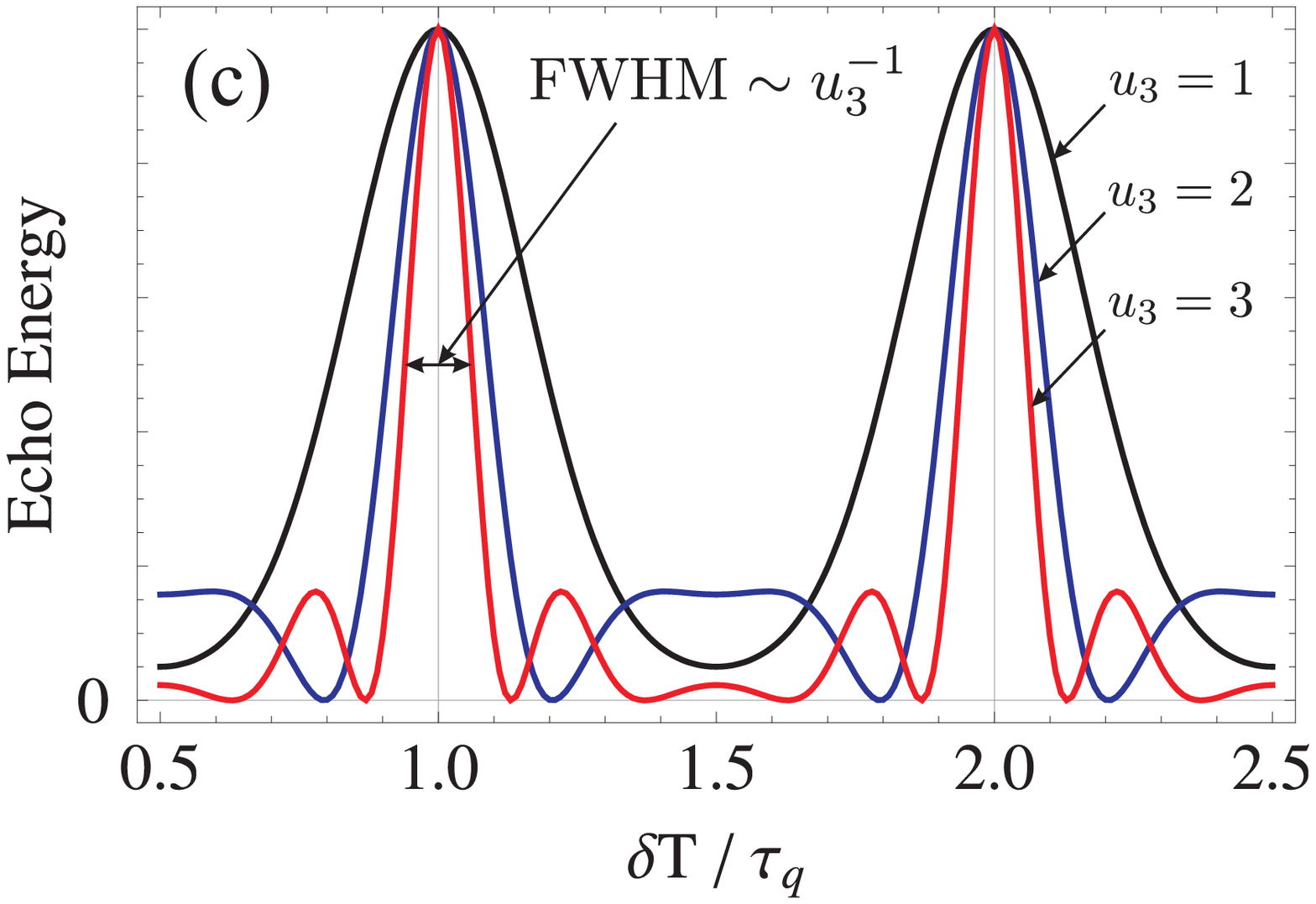}}
  \caption{(Color online) (a) An example of two, low-order trajectories that contribute to the two-pulse AI (SW$j$ = $j^{\rm{th}}$ sw pulse, RO = read-out pulse). Three momentum states are shown ($\ket{p_0}$, $\ket{p_0 + \hbar q}$ and $\ket{p_0 + 2\hbar q}$) corresponding to the solid, dashed and dotted lines, respectively. The sum of the Doppler and recoil phases are indicated for each of the two states that interfere at $t = 2T$. The Doppler phase difference, $q v_0 (t - 2T)$, is zero at the echo time for all initial atomic velocities, $v_0 = p_0/M$. The remaining phase difference, $\omega_q (3 t - 4 T)$, is a result of atomic recoil, and is equal to $2\omega_q T$ at the echo time. (b) Low-order trajectories for the three-pulse perturbative AI. Here, a third sw pulse, SW3, is applied to perturb the phase of the interference at $2T$. The Doppler phase difference is zero at $t = 2T$ \emph{and} independent of $\delta T$ for only those trajectories that differ by $\hbar q$ after SW3 \emph{and} SW2. (c) Echo energy as a function of $\delta T / \tau_q$ for the perturbative AI, where $\tau_q = \pi/\omega_q$ is the recoil period. Line shapes are shown for three different pulse areas, $u_3$, to illustrate the effect of fringe narrowing that occurs for increasing interaction strength. Here, we assume that only one ground state magnetic sub-level contributes to the signal.}
  \label{fig:1-AITheory}
\end{figure*}

The focus of this work is a precise determination of $\alpha$ from the ratio $h/M_{\rm{Rb}}$ using a three-pulse, perturbative grating-echo AI. This type of interferometer has been described in previous work \cite{Beattie-PRA(R)-2009} using the concept of coherence functions, and in \Refs \cite{Beattie-PRA-2009, Barrett-Thesis-2012} using a quantum-mechanical treatment. Experiments that utilize a similar multi-pulse interferometer for quantum chaos and kicked rotor studies can be found in \Refs \cite{Wu-PRL-2009, Tonyushkin-PRA(R)-2009}. Other work using the grating-echo AI is reviewed in \Ref \cite{Barrett-Advances-2011}.

The remainder of this article is organized as follows. In \Sec \ref{sec:Theory}, we provide a description of the interferometer. Section \ref{sec:Experiment} briefly discusses the setup of the experiment. Our primary results are given in \Sec \ref{sec:Results}, which is followed by \Sec \ref{sec:Conclusion} with a discussion regarding systematic effects and future work.

\section{Description of the AI}
\label{sec:Theory}

The grating-echo AI is a time-domain Talbot-Lau interferometer \cite{Clauser-PRA(R)-1994, Chapman-PRA(R)-1995, Berman-Book-2011}, the principles of which can be understood on the basis of a plane-wave description of the two-pulse scheme shown in \Fig \ref{fig:1-AITheory}(a) \cite{Cahn-PRL-1997, Strekalov-PRA-2002, Beattie-PRA-2008, Barrett-PRA-2010, Barrett-Advances-2011, Barrett-PRA-2011}. The AI relies on matter-wave interference produced by Kapitza-Dirac scattering of atoms by short, off-resonant sw pulses. Two sw pulses, spaced in time by $T$, are applied to a sub-Doppler laser-cooled sample with a root-mean-squared (rms) momentum spread of $p_{\rm{rms}} = (M k_B \mathcal{T})^{1/2} \gg \hbar k$, where $\mathcal{T}$ is the sample temperature and $k_B$ is Boltzmann's constant. For each atom in the sample, the first pulse excites a superposition of momentum states separated by integer multiples, $m$, of $\hbar q$. The second excitation pulse further diffracts the momentum states, causing certain trajectories to interfere in the vicinity of $t = 2T$, which we call the ``echo'' time. This interference creates a spatial modulation in the probability density for any given atom.

Between the pulses, the wave function associated with each momentum state, $\ket{p = p_0 + m\hbar q}$, evolves with a time-dependent phase
\be
  \label{eqn:phi}
  \phi = \frac{(p_0 + m\hbar q)^2}{2M} \frac{t}{\hbar} = \phi_0 + \phi_D + \phi_q
\ee
due to its kinetic energy. This phase has three contributions: the initial phase, $\phi_0 = (p_0^2/2M)t/\hbar$, due to the initial momentum $p_0$ of the atom at the time of the first pulse; the Doppler phase, $\phi_D = m q v_0 t$, where $v_0 = p_0/M$; and the recoil phase, $\phi_q = m^2 \omega_q t$. $\phi_0$ is unimportant for interference because it is the same for all excited momentum states. At $t = 2T$, the contribution to the interference due to the Doppler shift of the moving atom, $\phi_D$, is equal for any two overlapping trajectories. Thus, in a manner reminiscent of photon echoes \cite{Abella-PR-1965}, the Doppler phase cancels between interfering momentum states for \emph{all} initial velocity classes. This results in a macroscopic density grating in the sample at the echo time. As time elapses, these momentum states dephase due to the distribution of initial velocities. Consequently, the echo has a finite coherence time given by: $\tau_{\rm{coh}} \sim \big(|m - m'| q v_{\rm{rms}} \big)^{-1}$, which is $\tau_{\rm{coh}} \sim 2$ $\mu$s for typical experimental conditions. Here, $m$ and $m' \neq m$ are integers representing separate interfering momentum states, $\ket{p_0 + m\hbar k}$ and $\ket{p_0 + m'\hbar q}$, and $v_{\rm{rms}} = p_{\rm{rms}}/M$ is the rms velocity of the sample. The remaining phase component of each momentum state, $\phi_q$, is due to the recoil of the atom after $m$ two-photon scattering events induced by the sw excitation field. This phase determines the contrast of the interference pattern at $t = 2T$. An example of two interfering trajectories is shown in \Fig \ref{fig:1-AITheory}(a), which correspond to $m = 1$ and $m' = 2$. The contrast of the interference pattern produced by these trajectories oscillates at a frequency $2\omega_q$ as a function of $T$. The macroscopic density grating, however, has contributions from all pairs of interfering momentum states from all excited atoms \cite{Cahn-PRL-1997}, where each pair contributes a different harmonic of $\omega_q$ to the contrast oscillation.

The contrast of the macroscopic grating is measured by applying a traveling-wave read-out pulse and detecting the intensity of the coherently Bragg-scattered light in the backward direction. Due to the nature of Bragg diffraction, this detection technique is sensitive to only the spatial harmonics of the density modulation that have a period equal to an integer multiple of $\lambda/2$, where $\lambda$ is the wavelength of the read-out light. In the plane-wave picture, only interfering momentum states that differ by $\hbar q$ can produce such a modulation. Thus, the interferometer is sensitive to only the fundamental spatial frequency of the grating, $q$, which produces a temporal modulation in the grating contrast that oscillates at $2\omega_q$. The time-integrated power of the back-scattered light (referred to as the echo energy) is a measure of the contrast produced by this interference. Experiments utilizing the two-pulse AI, where the echo energy is measured as a function of $T$, are described in \Refs \cite{Cahn-PRL-1997, Beattie-PRA-2008, Barrett-PRA-2010, Barrett-Advances-2011, Barrett-PRA-2011, Barrett-Thesis-2012}.

For the three-pulse perturbative AI, an additional sw pulse is applied between the first two pulses at $t = \delta T < T$, as shown in \Fig \ref{fig:1-AITheory}(b) \cite{Beattie-PRA(R)-2009, Beattie-PRA-2009, Wu-PRL-2009, Tonyushkin-PRA(R)-2009, Barrett-Thesis-2012}. This pulse has the effect of diffracting the atom into higher-order momentum states that contribute additional harmonics of $\omega_q$ to the temporal modulation of the grating contrast. An example of a pair of interfering trajectories created by the three-pulse AI is shown in \Fig \ref{fig:1-AITheory}(b) \footnote{We emphasize that only a small subset of the trajectories excited by the sw pulses will interfere at $t = 2T$ for an arbitrary third pulse time, $\delta T$ (\ie trajectories which, when combined, exhibit a Doppler phase that is independent of $\delta T$). Specifically, the only momentum states contributing to the signal are those that differ by $\hbar q$ after SW3 \emph{and} after SW2.}. The resulting signal consists of a series of narrow fringes separated by the recoil period, $\tau_q = \pi/\omega_q$ ($\sim 32$ $\mu$s for rubidium), as a result of the interference between all excited momentum states that differ by $\hbar q$. Intuitively, the action of the third sw pulse is to perturb the phase of the momentum states by $\eta \omega_q \delta T$, where $\eta$ is an integer that depends on the particular pathways that lead to interference at $t = 2T$. Thus, as a function of $\delta T$, the contrast of the interference undergoes periodic revivals analogous to a multi-slit experiment in classical optics.

When all relevant trajectories are summed over, it can be shown \cite{Beattie-PRA-2009, Barrett-Thesis-2012} that the resulting echo energy is modulated by $J_0[2 u_3 \sin(\omega_q \delta T)]^2$, provided the third pulse area, $u_3$, is small (\ie $u_3 = \Omega_0^2 \tau_3/2|\Delta| \lesssim 1$). Here, $J_0(x)$ is the zeroth-order Bessel function of the first kind, $\Omega_0$ is the one-photon Rabi frequency, $\tau_3$ is the third sw pulse duration, and $\Delta$ is the detuning from the excited state. Figure \ref{fig:1-AITheory}(c) illustrates the predicted dependence of the echo energy (for a \emph{single} ground state magnetic sub-level) as a function of $\delta T$. The sensitivity of this AI to $\omega_q$ scales inversely with the time scale ($T$) over which the signal can be measured, and it scales proportionately with the width of the fringes. The advantage of using this AI over the two-pulse configuration is the ability to narrow the fringe width with the parameters of the third pulse. Additionally, since $T$ is fixed, the same number of atoms remain in the excitation beams at the time of detection---thus, there is no signal decay as a function of $\delta T$ due to time-dependent effects like the thermal expansion of the sample. The fringe width is effectively determined by the width of the excited momentum distribution. By increasing the proportion of high-order momentum states (and thus the higher harmonics of $\omega_q$), that contribute to the signal, the fringes become more sharply defined. The excitation is controlled by the interaction strength and duration of the third sw pulse. It can be shown that, for small pulse durations [\ie $\tau_3 \ll (|\Delta|/\Omega_0^2\omega_k)^{1/2}$, where $\omega_k = \hbar k^2/2M$], the full-width at half-maximum (FWHM) of the signal scales inversely with $u_3$ \cite{Beattie-PRA-2009, Barrett-Thesis-2012}. This feature is also illustrated in \Fig \ref{fig:1-AITheory}(c).

\section{Experimental Setup}
\label{sec:Experiment}

As described in \Refs \cite{Barrett-PRA-2011,Barrett-Thesis-2012}, two major improvements to the grating-echo AI experiment have enabled us to reach time scales of $T \sim 50$ ms: (i) utilizing a non-magnetizable glass vacuum system, which reduced decoherence effects related to inhomogeneos $B$-fields and improved the molasses cooling of the sample, and (ii) using large-diameter, chirped excitation beams, which eliminated the differential Doppler shift of the atom from the two components of the sw pulses and increased the transit time of the atoms in the beam. Magnetic field curvature produced by a stainless-steel vacuum chamber, and the gravity-induced, differential Doppler shift limited previous experiments to $T \lesssim 10$ ms \cite{Beattie-PRA-2008, Beattie-PRA(R)-2009, Beattie-PRA-2009}.

This experiment utilizes a laser-cooled sample of rubidium typically containing $\sim 5 \times 10^9$ atoms at temperatures of $\mathcal{T} \lesssim 5$ $\mu$K. Either $^{85}$Rb or $^{87}$Rb atoms are loaded into a magneto-optical trap (MOT) from a background Rb vapor. Prior to the AI experiment, the sample is prepared in the upper hyperfine atomic ground state (5S$_{1/2}$ $F = 3$ for $^{85}$Rb or $F = 2$ for $^{87}$Rb). The light for the AI is derived from a Ti:sapphire laser (linewidth $\sim 1$ MHz) that is locked above the D2 cycling transition using Doppler-free saturated absorption spectroscopy. A network of acousto-optic modulators (AOMs) is used to generate the frequencies necessary for the AI excitation and the read-out beams. The read-out light is detuned by $\Delta_{\rm{RO}} = 40$ MHz from the cycling transition---a condition that increases the back-scattered light intensity from the atoms \cite{Barrett-Thesis-2012}. The AI beams are detuned by $\Delta_{\rm{AI}} = 220$ MHz, and a frequency chirp of $\delta(t) = g t/\lambda$ is added to (subtracted from) the downward traveling (upward traveling) component of the sw pulses such that each excitation is kept on resonance with the two-photon transition as the sample falls in gravity \cite{Barrett-PRA-2011,Barrett-Thesis-2012}. Here, $g$ is the gravitational acceleration, and $\lambda$ is the wavelength of the AI light. A ``gate'' AOM is used upstream of the AI AOMs as a frequency shifter and as a high-speed shutter to reduce the amount of stray light in the experiment. All rf sources and digital-delay generators used to define the pulse timing for the AI are externally referenced to a 10 MHz rubidium clock.

\begin{figure}[!tb]
  \centering
  \includegraphics[width=0.48\textwidth]{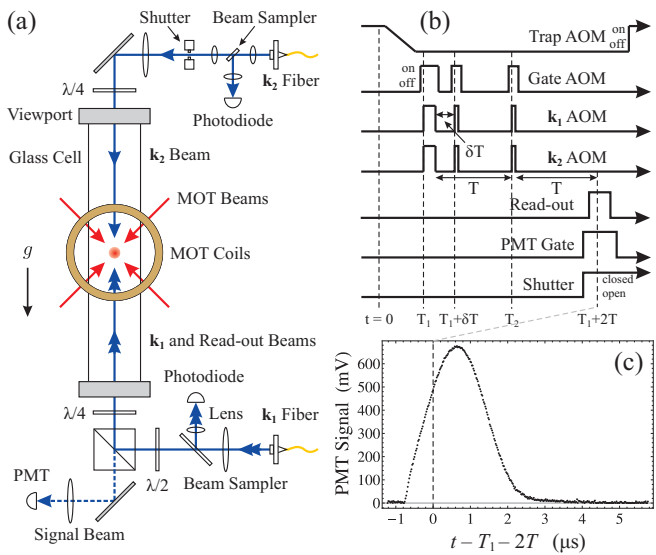}
  \caption{(Color online) (a) Optical setup for the interferometer. The glass cell has dimensions $7.6 \times 7.6 \times 84$ cm and is oriented along the vertical. (b) Timing diagram for the AI. The gate AOM is pulsed on to allow light for each excitation pulse produced by the $k_1$ and $k_2$ AOMs. The pulse occurring at $t = T_1 + \delta T$ corresponds to the perturbative sw pulse. The read-out pulse (which is independent of the gate AOM) and the PMT gate are turned on for $\sim 9$ $\mu$s in the vicinity of the echo time, $t = T_1 + 2T$. (c) Example of a two-pulse grating-echo signal (from a 10 $\mu$K $^{87}$Rb sample) recorded by the PMT, which corresponds to an echo energy of 130 pJ. AI pulse spacing: $T = 1.06338$ ms; pulse durations: $\tau_1 = 3.8$ $\mu$s, $\tau_2 = 1.2$ $\mu$s; AI and read-out beam detunings: $\Delta_{\rm{AI}} = 220$ MHz, $\Delta_{\rm{RO}} = 40$ MHz; AI and read-out beam intensity: $I \sim 40$ mW/cm$^2$.}
  \label{fig:2-Experiment}
\end{figure}

The AI beams are coupled into two AR-coated, single-mode optical fibers and aligned through the sample, as shown in \Fig \ref{fig:2-Experiment}(a). At the output of the fibers, the beams are expanded to a $e^{-2}$ diameter of $d \sim 1.7$ cm and are circularly polarized (in the $\sigma^+$-$\sigma^+$ or the $\sigma^-$-$\sigma^-$ configuration) by a pair of $\lambda/4$ wave plates. The timing sequence for the experiment is illustrated in \Fig \ref{fig:2-Experiment}(b). A mechanical shutter on the upper platform closes before the read-out pulse in order to block the back-scatter of read-out light produced by various optical elements. This light would otherwise interfere with the coherent signal from the atoms. A gated photo-multiplier tube (PMT) is used to detect the power in the back-scattered field. \Fig \ref{fig:2-Experiment}(c) shows an example of the echo signal from the two-pulse AI.

\section{Results}
\label{sec:Results}

\begin{figure*}[!tb]
  \centering
  \includegraphics[width=0.96\textwidth]{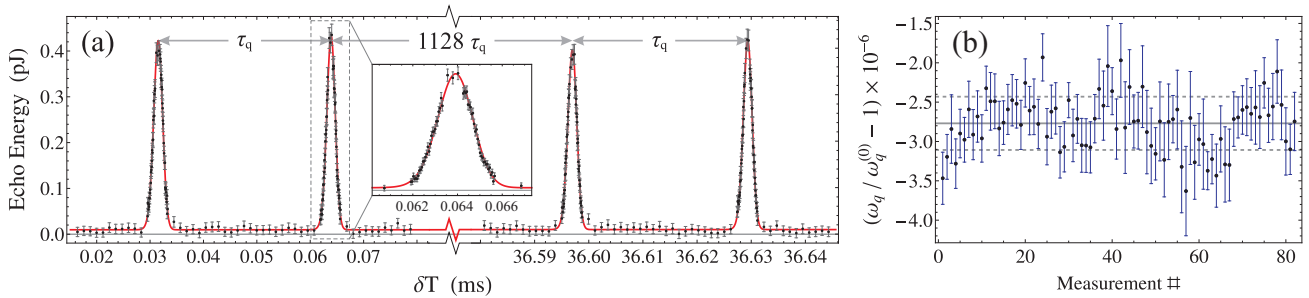}
  \caption{(Color online) (a) Demonstration of an individual recoil measurement in $^{85}$Rb using the perturbative AI at $T = 36.6656$ ms. Data are recorded in two temporal windows separated by $1128\,\tau_q \sim 36.5$ ms. The relative statistical uncertainty in $\omega_q$ is $\sim 180$ ppb, as determined from a least-squares fit. Inset: expanded view of the fringe near $\delta T = 64$ $\mu$s. (b) 82 independent measurements of $\omega_q$ in $^{87}$Rb displayed in chronological order. Each data point was recorded in $\sim 10$ minutes of data acquisition time, with a typical statistical uncertainty of $\sim 380$ ppb. Measurements are scaled by the expected value of the recoil frequency, $\omega_q^{(0)} = 94.77384783(12)$ rad/ms, which is based on the value of $h/M$($^{87}$Rb) from \Ref \cite{Bouchendira-PRL-2011} and the $F = 2 \to F' = 3$ transition frequency in $^{87}$Rb from \Ref \cite{Steck-Rb87Data-2010}. The dashed grid lines indicate the weighted standard deviation of 339 ppb, and the standard deviation of the mean is 37 ppb. The corresponding reduced chi-squared is $\chi^2/\mbox{dof} = 0.93$ for $\mbox{dof} = 81$ degrees of freedom. The mean value, shown by the solid grid line, is $\sim 2.8$ ppm below the expected value, which is due to systematic effects. AI pulse parameters: $T = 45.4837$ ms, $\tau_1 = 2.2$ $\mu$s, $\tau_2 = 1.4$ $\mu$s, $\tau_3 = 3$ $\mu$s, $\Delta_{\rm{AI}} = 219.8$ MHz, $\Delta_{\rm{RO}} \sim 40$ MHz, $I \sim 95$ mW/cm$^2$.}
  \label{fig:3-Results}
\end{figure*}

Measurements of $\omega_q$ were obtained using the perturbative three-pulse AI by measuring the echo energy as a function of the third pulse time, $\delta T$, as shown in \Fig \ref{fig:3-Results}(a). This figure shows a measurement of $\omega_q$ in $^{85}$Rb on a time scale of $T \sim 36.7$ ms, which was acquired in $\sim 15$ minutes. Clearly, the shape of the fringes does not resemble that predicted by the theory shown in \Fig \ref{fig:1-AITheory}(c). This is due to the contribution from each of the magnetic sub-levels in the $F = 3$ ground state of $^{85}$Rb, which tend to smear out the higher harmonics in the signal---a result of the different coupling strengths of these states. Furthermore, the presence of additional, nearby excited states ($F' = 2$ and 3 in the case of $^{85}$Rb) has been shown to produce an asymmetry in the fringe line shape \cite{Barrett-Thesis-2012}. This effect is reduced in $^{87}$Rb because the frequency difference between neighboring excited states is larger. To measure $\omega_q$, the data are fit to a phenomenological model that consists of a periodic sum of exponentially-modified Gaussian functions:
\begin{align}
\begin{split}
  F(\delta T; \tau_q)
  & = \sum_l A_l \, \exp\left[\frac{1}{2} \left(\frac{\sigma_l}{\upsilon}\right)^2 + \frac{\delta T - l \tau_q}{\upsilon} \right] \\
  & \times \mbox{erfc} \left[ \frac{1}{\sqrt{2}} \left( \frac{\sigma_l}{\upsilon} + \frac{\delta T - l \tau_q}{\sigma_l} \right) \right],
\end{split}
\end{align}
and the recoil frequency, $\omega_q = \pi/\tau_q$, is extracted from the fit. In this model, erfc$(x)$ is the complementary error function, and the parameter $\upsilon$, which determines the amount of asymmetry in the line shape, is the same for all fringes. The fit to the data shown in \Fig \ref{fig:3-Results}(a) yielded a reduced chi-squared of $\chi^2/\mbox{dof} = 0.51$ for $\mbox{dof} = 300$ degrees of freedom. This corresponds to a relative statistical precision of $\sim 180$ ppb in $\omega_q$---representing a factor of $\sim 9$ improvement over previous work \cite{Beattie-PRA(R)-2009}.

To demonstrate the statistical uncertainty of the measurement under current conditions, 82 independent measurements of $\omega_q$ in $^{87}$Rb were obtained (with all other experimental conditions held fixed to the extent possible). The distribution of individual recoil measurements is shown in \Fig \ref{fig:3-Results}(b). Here, $\omega_q$ is determined from a weighted average over all individual measurements, where the points are weighted inversely proportional to the square of their statistical uncertainties. The mean value shown in the figure, which has not been corrected for systematic effects, is found with a relative statistical uncertainty of 37 ppb, as determined by the standard deviation of the mean.

An autocorrelation analysis of these measurements indicates that the results are correlated at the 20\% level with measurements taken at a previous time. This is attributed to slowly varying lab conditions over the 14 hours of data acquisition time. The primary contributors to these correlations are the temperature and the time-varying ambient magnetic environment of the lab, which are currently being stabilized for a new round of measurements.

\subsection{Systematic Effects}

We have investigated systematic effects on the measurement of $\omega_q$ related to the angle between excitation beams, the refractive indices of the sample and the background Rb vapor, light shifts, Zeeman shifts, $B$-field curvature and the sw pulse durations \cite{Barrett-Thesis-2012}. The total systematic uncertainty in this measurement is estimated to be $\sim 5.7$ ppm, and is dominated by two effects: (i) the refractive index of the sample, and (ii) the curvature of the $B$-field that the atoms experience as they fall under gravity. We now discuss these two effects in detail.

The refractive index of the atomic sample affects the wave vector of the excitation beams, since a photon in a dispersive medium acts as if it has momentum $n\hbar k$, where $n$ is the index of refraction \cite{Campbell-PRL-2005}. For near-resonant light, the index becomes a function of both the density of the medium, $\rho$, and the detuning of the applied light from the atomic resonance, $\Delta_{\rm{AI}}$. The systematic effect on the recoil frequency due to the refractive index can be expressed as $\omega_q(\rho,\Delta_{\rm{AI}}) = \omega_q^{(0)} n^2(\rho,\Delta_{\rm{AI}})$, where $\omega_q^{(0)}$ is the recoil frequency in the absence of systematic effects. The index of refraction can be computed from the electric susceptibility and the light-induced polarization of the medium \cite{Campbell-PRL-2005}. Taking into account the level structure of the atom, it can be shown that \cite{Barrett-Thesis-2012}
\be
  \label{eqn:n(rho,DeltaHG)}
  n(\rho,\Delta_{HG}) = \sqrt{1 - \frac{\rho}{\epsilon_0 \hbar \Gamma} \sum_{H} \mu_{HG}^2 \frac{\Delta_{HG}/\Gamma}{1 + (\Delta_{HG}/\Gamma)^2}}.
\ee
Here, $\Delta_{HG} \equiv \omega - (\omega_H - \omega_G)$ is the atom-field detuning between the ground and excited manifolds, $\ket{g,G}$ and $\ket{e,H}$, for laser frequency $\omega$. $G$ ($H$) is a quantum number representing the total angular momentum of a particular ground (excited) manifold, and $\mu_{HG}$ is the reduced dipole matrix element for transitions between those manifolds \cite{Berman-Book-2011}.

There are two separate sources of correction due to the index of refraction in our experiment: the background vapor of rubidium, and the sample of cold atoms. However, since the density of background vapor is typically two orders of magnitude less than the rms density of the trap, the systematic correction to $\omega_q$ is dominated by the cold atoms. Nevertheless, the correction due to the background vapor is non-negligible ($-140$ ppb for a background density of $\sim 5 \times 10^8$ atoms/cm$^3$ and $\Delta_{\rm{AI}} = 220$ MHz). For the MOT, the rms density at the time of trap release was measured to be $4.1(1.2) \times 10^{10}$ atom/cm$^3$ based on time-of-flight images \cite{Barrett-Thesis-2012}. We estimate a shift in $\omega_q$ of $-10.5(3.0)$ ppm at a detuning of $\Delta_{\rm{AI}} = 220$ MHz. We discuss how this systematic can be addressed in \Sec \ref{sec:Conclusion}.

The other dominating systematic effect is due to the inhomogeneity of the magnetic field sampled by the atoms during the interrogation time of the interferometer ($2T \sim 100$ ms). This field primarily originates from nearby ferromagnetic material, such as an ion pump magnet and a glass-to-metal adaptor, and from the set of quadrupole coils we use to cancel the residual field in the vicinity of the MOT \cite{Barrett-PRA-2011, Barrett-Thesis-2012}. To quantify this effect, we have carried out a calculation similar to that shown in \Ref \cite{Barrett-PRA-2011} where, instead of a field that varies linearly in space, the local $B$-field along the vertical direction is modeled by $B_z(z) = \beta_0 + \beta_1 z + \beta_2 z^2/2$. Here, the quantities $\beta_0$, $\beta_1$, and $\beta_2$ are constant, and the curvature is assumed to be small such that $\beta_0 \sim \beta_1 z \sim \beta_2 z^2/2$ are all comparable over the length scale of the interferometer, $z \sim 5$ cm. The constant term in this model, $\beta_0$, gives rise to a Zeeman shift in each magnetic sub-level---the effects of which are negligible on this measurement ($\sim 0.15$ ppb for 20 mG of residual $B$-field). In previous work \cite{Barrett-PRA-2011}, we showed that there is no systematic effect on $\omega_q$ due to $\beta_1$. Since a constant $B$-gradient gives rise to a constant force on the atoms, the sole effect of $\beta_1$ is to phase shift the gratings associated with each magnetic sub-level---similar to the effects of gravity. There is no shift in the measurement of $\omega_q$ due to $\beta_1$ because the force acts equally on all pathways of the interferometer. However, the curvature term, $\beta_2$, is responsible for a position-dependent force similar to a harmonic oscillator. Thus, for each momentum state trajectory, the atom samples a different region of space and experiences a different acceleration than that of a neighboring trajectory. This picture explains how a curved $B$-field can affect a measurement of $\omega_q$, since the momentum of each trajectory is differentially modified between excitation pulses.

The contribution to the recoil phase due to the $B$-field curvature is proportional to $g_F m_F \mu_B \beta_2 \omega_q T^3/M$, where $g_F$ is a g-factor, $m_F$ is the magnetic quantum number of the ground state, and $\mu_B$ is the Bohr magneton. This implies that, even for a small curvature ($\sim 1$ mG/cm$^2$), the phase shift can be significant ($\sim 1$ rad) for moderate pulse spacings ($T \sim 50$ ms). It also explains how small inhomogeneities in the field can lead to significant decoherence effects and therefore a decrease in signal lifetime. Under current experimental conditions, the signal is expected to be dominated by the extreme state $\ket{F,m_F=F}$, where the phase shift is the largest among all sub-levels. For a constant $B$-field curvature, the corresponding systematic correction to $\omega_q$ is
\be
  \omega_q(\beta_2,T) = \omega_q^{(0)} \left[ 1 + \frac{2}{3} \left( \frac{m_F g_F \mu_B \beta_2}{M} \right) T^2 \right].
\ee
Assuming $|\beta_2| \sim 0.1$ mG/cm$^2 = 10^{-4}$ T/m$^2$ (an estimate based on measurements of the $B$-field in the vicinity of the MOT using a flux-gate sensor), and a time scale of $T = 50$ ms, the relative correction to $\omega_q$ is $\sim \pm 11$ ppm for the $\ket{F = 2, m_F = 2}$ state in $^{87}$Rb, where the sign of the shift depends on the sign of $\beta_2$. This effect clearly results in a significant shift in $\omega_q$.

However, since neither the spatially varying $B$-field that the atoms experience, nor the distribution of atoms in the ground state magnetic sub-levels, is well known, we use a measurement of the variation in $\omega_q$ as a function of $T$ to estimate the shift due to this systematic. In using this method, we implicitly account for all systematic shifts in $\omega_q$ that vary with $T$.

\begin{figure}[!tb]
  \centering
  \includegraphics[width=0.40\textwidth]{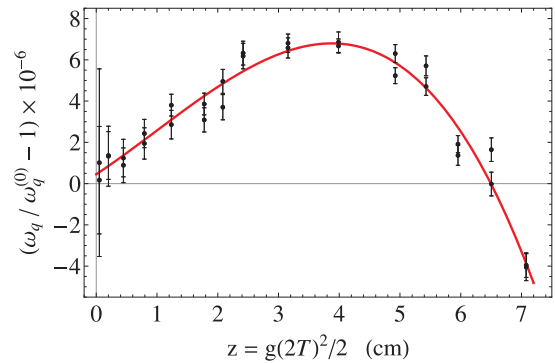}
  \caption{(Color online) Measurements of $\omega_q$ as a function of $z = g(2T)^2/2$. The vertical axis is scaled by the expected value of $\omega_q^{(0)} = 94.77384783$ rad/ms for excitation light at $\Delta_{\rm{AI}} = 219.8$ MHz above the $F = 2 \to F' = 3$ transition in $^{87}$Rb. The variation in $\omega_q$ spans roughly 12 ppm, which is attributed to a combination of the spatially varying $B$-field curvature and the time-varying refractive index of the cloud. The canceling $B$-fields were set to achieve the largest signal lifetime, which in this case is $2T \sim 120$ ms. Measurements of $\omega_q$ at each $T$ were taken at random. Repeated measurements at the same $T$ are mostly consistent, but show a slight variation outside the statistical uncertainty indicated by the error bars. This is attributed to instability in the magnetic field environment of the lab. The red curve is a fit to a third-order polynomial. AI pulse parameters: $\tau_1 = 2$ $\mu$s, $\tau_2 = 1.4$ $\mu$s, $\tau_3 = 2$ $\mu$s, $\Delta_{\rm{AI}} \sim 220$ MHz, $\Delta_{\rm{RO}} \sim 40$ MHz, $I \sim 45$ mW/cm$^2$.}
  \label{fig:4-Systematic-BCurvature}
\end{figure}

Figure \ref{fig:4-Systematic-BCurvature} shows measurements of $\omega_q$ as a function of the center-of-mass position of the cloud, $z = g(2T)^2/2$, for various $T$. These data were taken under the same conditions as the measurement shown in \Fig \ref{fig:3-Results}(b), and clearly indicate the presence of a systematic shift in $\omega_q$ as $T$ changes. This is attributed to mechanisms that affect $\omega_q$ as a function of $t$, where $t = 0$ represents the release time of the trap. Two examples of such mechanisms are the spatially varying $B$-field that the atoms experience as they drop under gravity, and the time-varying refractive index of the thermally expanding atomic cloud. A separate observation, where a small change in the canceling coil currents produced a shift in the echo time for large $T$, indicates that the recoil phase, $\phi_q$, can be modified by magnetic effects. These data provide convincing evidence that a $B$-field curvature contributes to the overall systematic shift of $\omega_q$.

For the purpose of determining the systematic shift on $\omega_q$ due to all time-varying effects, we have fit the data shown in \Fig \ref{fig:4-Systematic-BCurvature} to a third-order polynomial. With this method, the determination of the appropriate shift amounts to finding the difference between the vertical offset of the fit function, and the value of the function corresponding to a pulse spacing of $T = 45.4837$ ms (or $z = 4.0575$ cm). In the absence of any $t$-dependent systematic effects, there is no variation in $\omega_q$ with $T$, and this offset will be zero. We estimate the shift due to all time-varying systematics to be $+6.3(4.4)$ ppm. This shift is thought to be dominated by the $B$-field curvature, since a separate estimate of the shift due to the time-varying refractive index of the sample gives approximately $+2$ ppm \cite{Barrett-Thesis-2012}.

\section{Discussion and Conclusion}
\label{sec:Conclusion}

In this section, we discuss techniques for reducing the aforementioned systematic effects. To reach competitive levels of measurement uncertainty with this interferometer requires a reduction in the systematic error by three orders of magnitude---presenting a significant challenge.

At first glance, \Eq \refeqn{eqn:n(rho,DeltaHG)} for the refractive index suggests that the relative correction to $\omega_q$ can only be reduced by decreasing the sample density, $\rho$, or by increasing the excitation beam detuning, $\Delta_{\rm{AI}}$. However, the current configuration of the AI relies on a large number of atoms to achieve a sufficient signal-to-noise ratio. Thus, a decrease in the sample density leads to a reduction in the signal size. Furthermore, the sensitivity of the three-pulse perturbative AI relies on a relatively strong atom-field coupling in order to excite many orders of momentum states. An increase in the excitation beam detuning without a corresponding increase in the field intensity leads to a reduction in the sensitivity of the AI to $\omega_q$. A $10^3$ reduction in this systematic could be accomplished by decreasing the rms density of the sample by a factor of 10, accompanied by a factor of 100 increase in the detuning. This would require an increase in the excitation field intensity by a factor of 100 (corresponding to $\sim 10$ W/cm$^2$) in order to retain the same sensitivity to $\omega_q$.

\begin{figure}[!tb]
  \centering
  \includegraphics[width=0.40\textwidth]{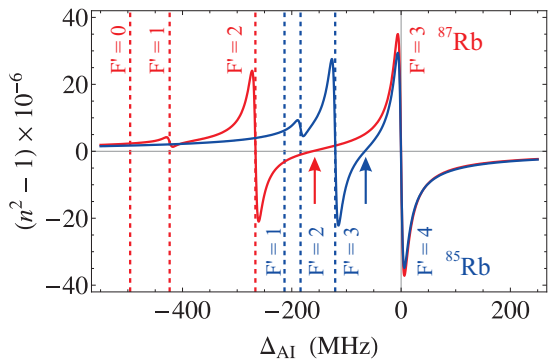}
  \caption{(Color online) Relative correction to the recoil frequency due to the refractive index as a function of the detuning of the excitation field, $\Delta_{\rm{AI}}$. These curves are based on \Eq \refeqn{eqn:n(rho,DeltaHG)} with a density of $\rho = 10^{10}$ atoms/cm$^3$. Predictions for both $^{85}$Rb and $^{87}$Rb are shown. The detuning is plotted with respect to the $F = 3 \to F' = 4$ transition in $^{85}$Rb, and the $F = 2 \to F' = 3$ transition in $^{87}$Rb. The dashed grid lines label the location of excited states \cite{Steck-Rb85Data-2010,Steck-Rb87Data-2010}. The ``magic'' frequencies, where the relative correction crosses zero, are indicated with arrows at $\Delta_{\rm{AI}} \approx -66.4$ MHz for $^{85}$Rb and at $\Delta_{\rm{AI}} \approx -162.6$ MHz for $^{87}$Rb.}
  \label{fig:5-Systematic-Detuning}
\end{figure}

A closer examination of the frequency-dependence of the refractive index reveals that there is a ``magic'' detuning where the relative shift in the recoil frequency ($n^2 - 1$) is exactly zero, as shown in \Fig \ref{fig:5-Systematic-Detuning}. This frequency is located between two excited state manifolds, where the dispersive corrections to $n$ due to each state have the same magnitude but opposite signs. For $^{85}$Rb, this magic detuning is between the $F' = 3$ and $F' = 4$ states at $\Delta_{\rm{AI}} \approx -66.4$ MHz, and for $^{87}$Rb it is located between the $F' = 2$ and $F' = 3$ states at $\Delta_{\rm{AI}} \approx -162.6$ MHz, as shown in \Fig \ref{fig:5-Systematic-Detuning}. Unlike the zero-crossings in the shift that are located in the vicinity of the two most energetic excited states, these magic frequencies are off-resonance---which is beneficial for reducing incoherent transitions due to spontaneous emission during the excitation pulses. Furthermore, these frequencies depend on only the reduced dipole matrix elements, $\mu_{HG}$, and the relative detuning between excited states, $\Delta_{HG}$. Since the magic detuning is independent of density, it is ideal for canceling both static and time-dependent shifts in $\omega_q$ due to the density of the sample, as well as the background vapor. By utilizing this property of the index correction, it should be possible to account for this systematic without reducing the sample density or requiring a very intense excitation beam. However, further experimental studies must be carried out to investigate the effect of light shifts on the excited states, and the corresponding correction to the magic detunings.

The systematic shift due to the $B$-field curvature can be significantly reduced by selecting only the $m_F = 0$ atoms to participate in the experiment. Then, any systematics due to the $B$-field would originate from the second-order Zeeman effect which shifts the $m_F = 0$ sub-levels by an amount proportional to $B^2$. A standard way of selecting only these atoms is to first optically-pump all of the atoms into the lower hyperfine ground state. Then, by applying a bias magnetic field to lift the degeneracy of magnetic sub-levels, and using a microwave pulse tuned across the hyperfine splitting of the ground states ($\sim 6.835$ GHz in the case of $^{87}$Rb) to drive a $\pi$-transition, the population of the $m_F = 0$ state can be transferred to the upper hyperfine level. This can be followed with a unidirectional ``blast'' beam, tuned on the repumping transition, to remove the remaining atoms in the lower state. With this technique, one can retain $\sim 1/3$ of $^{87}$Rb atoms ($\sim 1/5$ of $^{85}$Rb atoms), but they are guaranteed to be in the magnetically insensitive $m_F = 0$ sub-level provided the microwave field is tuned correctly.

Utilizing only $m_F = 0$ atoms in the experiment will have the added benefit of significantly reducing decoherence due to the $B$-field curvature---enabling an increase in $T$ and a corresponding reduction in the statistical error of each measurement. Under current conditions, we have achieved a maximum time scale of $T \sim 65$ ms. However, previous studies indicate that the transit time of the atoms in the excitation beams is $\sim 270$ ms \cite{Barrett-PRA-2011}, suggesting that $T$ can be as large as $\sim 135$ ms before the temperature of the sample becomes the limiting factor. Furthermore, since the $\ket{F, m_F = 0} \to \ket{F' = F, m_{F'} = 0}$ transition is not allowed, the effects of the nearest-neighbor excited state ($F' = 2$ in $^{87}$Rb) on the line shape of the AI signal can be reduced if a linearly-polarized excitation beam is used.

It is also desirable to increase the signal-to-noise ratio in the experiment---a quantity that strongly affects the statistical uncertainty of the measurement. Using knowledge of the energy in the back-scattered signal, and the power of the read-out pulse, it is possible to estimate the reflection coefficient, $R$, of the grating-echo. We find $R \sim 0.001$ under typical experimental conditions. One method of increasing this quantity is by pre-loading the sample in an optical lattice such that the initial spatial distribution has a significant $\lambda/2$-periodic component \cite{Andersen-PRL-2009}. Experimental studies of MOTs loaded into an intense, off-resonant optical lattice have shown that the reflection coefficient of the light that is Bragg-scattered off the resulting atomic grating can be as large as $R \sim 0.8$ \cite{Schilke-PRL-2011}. This motivates the pursuit of high-contrast grating production using a far-detuned lattice pulse that precedes the AI excitations. Such an endeavor would require an apparatus with good stability and control of the phase of the sw fields to (i) effectively channel atoms into the nodes of the lattice potential, and (ii) to match the phases of the excitation and lattice fields. We do not anticipate any significant systematic effects to arise from the lattice field because it would not be part of the interferometer. However, further theoretical and experimental investigations must be carried out to confirm these expectations.

With these experimental improvements, we anticipate that a future round of measurements will yield results with both statistical and systematic uncertainties at competitive levels.

\section*{Acknowledgements}

This work was supported by the Canada Foundation for Innovation, Ontario Innovation Trust, the Natural Sciences and Engineering Research Council of Canada, Ontario Centres of Excellence, the US Army Research Office and York University. We thank Tycho Sleator of New York University for generously lending crucial components of a Ti:sapphire laser that enabled the completion of this work.

\bibliographystyle{apsrev4-1}
\bibliography{MasterBib}
\end{document}